# Real-time observation of interfering crystal electrons in high-harmonic generation

M. Hohenleutner[1], F. Langer[1], O. Schubert[1], M. Knorr[1], U. Huttner[2],

S. W. Koch[2], M. Kira[2†], and R. Huber[1†]

*1 Department of Physics, University of Regensburg, 93040 Regensburg, Germany*

*2 Department of Physics, University of Marburg, 35032 Marburg, Germany*

[†]Authors to whom correspondence should be addressed

**Accelerating and colliding particles has been a key strategy to explore the texture of matter. Strong lightwaves can control and recollide electronic wavepackets, generating high-harmonic (HH) radiation which encodes the structure and dynamics of atoms and molecules and lays the foundations of attosecond science[1-3]. The recent discovery of HH generation in bulk solids[4-6] combines the idea of ultrafast acceleration with complex condensed matter systems and sparks hope for compact solid-state attosecond sources[6-8] and electronics at optical frequencies[3,5,9,10]. Yet the underlying quantum motion has not been observable in real time. Here, we study HH generation in a bulk solid directly in the time-domain, revealing a new quality of strong-field excitations in the crystal. Unlike established atomic sources[1-3,9,11], our solid emits HH radiation as a sequence of subcycle bursts which coincide temporally with the field crests of one polarity of the driving terahertz waveform. We show that these features hallmark a novel non-perturbative quantum interference involving electrons from multiple valence bands. The results identify key mechanisms for future solid-state attosecond sources and next-generation lightwave electronics. The new quantum interference justifies the hope for all-optical bandstructure reconstruction and lays the foundation for possible quantum logic operations at optical clock rates.**

Ultrafast time resolution in the few-femtosecond or attosecond regime has provided systematic insight into quantum control of individual atoms[12], molecules[13], and solids[14]. A spectacular paradigm has been to utilize the carrier wave of strong light pulses to control subcycle electron motion in atoms and molecules and follow the wavepacket dynamics directly via the temporal structure of HH emission[1-3,11,15]. Quantum theories[16] suggest, e.g., that maximum HH emission occurs at a distinct delay after the crest of the driving field, reflecting the time needed to accelerate electrons in the continuum[16,17]. Subcycle resolution has also been used to unravel novel interference phenomena in molecules[15,18,19].

In comparison, subcycle control of electrons in solids is still in its infancy, despite its promise of novel quantum physics[4-7,10,20-23] and fascinating applications in all-optical bandstructure reconstruction[22,23], lightwave-driven electronics[3,5,9,10,24] or attosecond science[6-8]. Only recently high-harmonic generation (HHG) has been extended to bulk solids, setting bandwidth records in the terahertz-to-ultraviolet spectral window[4,5]. An intriguing interplay of coherent interband polarization and intraband electron acceleration in the regime of dynamical Bloch oscillations has been suggested to underlie HHG in bulk crystals[5,7,20-23], explaining e.g. the observed linear scaling of the HH cut-off frequency with the driving peak field[4,5] in contrast to a quadratic behaviour in atoms and molecules. A detailed understanding of the microscopic electron motion as well as all envisaged applications depend critically on a direct access to the temporal structure of HH from bulk crystals[6,23,25], which has been elusive.

Here, we resolve the temporal fine structure of terahertz-driven phase-locked HH pulses from a bulk semiconductor. In addition, we directly measure the HH timing with respect to the driving field on the same absolute timescale for the first time. Our data reveal that the radiation is emitted as a train of almost bandwidth-limited bursts synchronised with the maxima of the field. Differently from atoms, the bursts are emitted only during every second half-cycle. We show that these signatures originate from a new type of non-perturbative interband quantum interference involving electrons below the Fermi energy.

Multi-octave spanning HH pulses (Extended Data Fig. 1) are generated by focusing intense phase-stable multi-terahertz (THz) transients centred at a frequency of $\nu_{THz} = 33$ THz (Fig. 1a, black waveform) into a single-crystal of the semiconductor gallium selenide (GaSe). To analyse the HH pulses with subcycle resolution, we introduce a novel combination of cross-correlation frequency-resolved optical gating (XFROG) and electro-optic sampling (Fig. 1a). The generated HH and the THz driving field (red waveform) are superimposed with a delayed 8-fs near-infrared gate (blue waveform) and focused into a 10-µm-thin BBO crystal. Nonlinear frequency mixing simultaneously yields sum-frequency (SF) signals encoding the temporal structure of HH pulses as well as electro-optic traces of the THz driving waveform (see 'Experimental setup' in Methods). In this way, the relative timing of HH emission with respect to the THz field is determined with an uncertainty corresponding to a fraction $T/20 = 1.5$ fs of the oscillation period $T$ of the driving waveform (see 'Determination of the absolute timescale' in Methods and Extended Data Fig. 2).

Figure 1 compares the THz pump field (Fig. 1b, black curve) with the spectrally integrated (Fig. 1b, shaded curve) and the spectrally resolved (Fig. 1c, colour map) SF signal. A double-blind XFROG algorithm (see 'Double-blind XFROG algorithm' in Methods) allows us to retrieve the actual temporal envelopes and relative phases of both the gate and the HH pulses[26] from the SF data. The consistency of this analysis is confirmed by the excellent agreement between the measured and reconstructed two-dimensional spectrograms (Figs. 1c and d) and between the intensity envelope of the gate pulse retrieved from the spectrogram and an independent second harmonic FROG measurement, respectively (Extended Data Fig. 3). The retrieved time trace of the HH intensity $I_{HH}(t)$ contains spectral contributions from 50 to 315 THz (Extended Data Fig. 3). $I_{HH}(t)$ consists of a train of three ultrashort bursts (Fig. 1e, shaded curve) featuring three remarkable properties: (i) The maxima of $I_{HH}(t)$ and $E_{THz}(t)$ coincide within $\pm 2$ fs $= T/15$ (vertical dashed lines). This behaviour is in contrast to ballistic electron recollision models[17] where the maximum of $I_{HH}(t)$ is distinctly delayed with respect to the maximal driving field[3]. (ii) Unlike in atomic HHG, $I_{HH}(t)$ is suppressed by one order of magnitude for field maxima of negative polarity. (iii) The duration of the unipolar HH bursts is as short as 7 fs (FWHM of intensity), which corresponds to a single oscillation period of the fourth harmonic order. Such pulse widths are expected only if all frequency components within the smooth spectral envelope (Extended Data Fig. 3) generated during one half-cycle of the driving field are emitted almost simultaneously. This is indeed the case as can be seen in Figs. 1c and d, where all SF components peak roughly at the same delay time *t* (vertical broken line in Figs. 1c,d), suggesting, at most, a weak spectral chirp of the HH pulses.

The observed time structure implies a quasi-instantaneous and unipolar generation mechanism. In order to identify this key ingredient, we first reproduce $I_{HH}(t)$ by a full quantum theory[5,20] (see 'Quantum many-body model' in Methods) including intra- and interband dynamics with two conduction and three valence bands (Extended Data Fig. 4). Our calculation reproduces the experimentally observed behaviour of $I_{HH}(t)$ in great detail (Fig. 2a, red solid curve). In particular, the emission peaks within 2 fs about the positive field crest while it is strongly suppressed for negative field extrema. In contrast, recent models accounting for only two electronic bands have consistently predicted HH emission in a bipolar way[7,25] and have suggested analogies with atomic HHG[22] where mostly two classes of electronic states have been considered: the ground state and the continuum of ionized states. In a solid, however, the simultaneous interaction of each electron with many atoms of the crystal lattice forms a series of electronic bands. As soon as more than two bands are included, electrons may be excited through multiple paths inducing quantum interference.

Figure 2b illustrates a minimal model for this scenario accounting for two valence ($h_1$ and $h_2$) and one conduction band ($e_1$). Excitation of an electron from band $h_1$ to band $e_1$ may either proceed by multi-photon transitions directly between two bands $h_1 \rightarrow e_1$, or indirectly via an additional band, $h_1 \rightarrow h_2 \rightarrow e_1$. The THz pulse is far off either resonance, but it is sufficiently strong to generate non-perturbative excitations where electron populations change drastically on a subcycle scale. We show that such non-perturbative transitions tend to balance the respective weights of the excitation paths because the extremely strong field forces the electrons to oscillate between the non-resonantly coupled states (see 'Interference path efficiency' in Methods). Nonetheless, the excitation paths maintain their perturbatively assigned symmetry (see 'Strong-field quantum interference' in Methods), featuring an odd transition amplitude $s(-E_{THz}) = -s(E_{THz})$ with respect to the driving field for the direct excitation and an even amplitude $t(-E_{THz}) = t(E_{THz})$ for the indirect path $h_1 \rightarrow h_2 \rightarrow e_1$ (Fig. 2b, Extended Data

Fig. 5). A coherent superposition of both yields a total amplitude of $t(|E_{THz}|) +(-) s(|E_{THz}|)$ for positive (negative) $E_{THz}$, respectively (see 'Strong-field quantum interference' in Methods). Hence, the sign of the field controls the total outcome of HH transitions. Note that the transition $h_1 \rightarrow h_2$ connecting bands below the Fermi level is initially Pauli-blocked but strong excitation can significantly empty $h_2$, e.g., via the transition $h_2 \rightarrow e_1$, clearing the path $h_1 \rightarrow h_2 \rightarrow e_1$.

We test the viability of this concept by a systematic switch-off analysis using our 5-band computation that includes all relevant transitions. By artificially multiplying the dipole moment $d_{h1h2}$ between the hole bands $h_1$ and $h_2$ as well as all other similar terms with a coherent control factor $F_{cc}$, we eliminate the indirect paths needed for non-perturbative quantum interference. Figure 2a compares the intensity envelope, $I_{HH}(t)$, with (red solid line, $F_{cc} = 1$) and without (red dashed line, $F_{cc} = 0$) the indirect paths. Switching off the quantum interference, i.e., considering only direct transitions ($F_{cc} = 0$), produces bursts at positive and negative crests of the field. Interestingly, the bursts become delayed by roughly T/4 with respect to the field extrema, which is consistent with a delay expected in an atomic recollision model[3,17]. However, opening the indirect paths ($F_{cc} = 1$) makes the emission instantaneous with the driving field. More specifically, the interband coherence is directly driven by the field such that the quantum interference and the resulting HH emission are strongest during the presence of the electric field. Hence, the quantum interference synchronises (suppresses) the emission with the positive (negative) crests of the field. The destructive interference is not perfect, leaving small HH remnants at negative peak fields.

Figure 2c shows a contour plot of computed normalized $I_{HH}(t)$ traces as a function of $F_{cc}$ (unscaled representation in Extended Data Fig. 6). By gradually suppressing the coherent-control paths, emission appears as delayed bursts after each field maximum and minimum. Nevertheless, the transition is not smooth, but contains nontrivial oscillations and bifurcations. These features are caused by THz-induced band mixing, which modulates electronic populations and $I_{HH}(t)$ and underpins the non-perturbative character of the interband excitations.

Under the extremely non-resonant conditions of our experiment ($h\nu_{THz} < E_g/14$ with $E_g = 2.0$ eV being the band gap energy of GaSe), band-to-band transitions require multiple THz pump photons. Non-perturbative excitations can non-resonantly drive all these transitions to exhibit large population transfer (Extended Data Fig. 5) and the related processes are robust against variations of the THz field strength and photon energy. Therefore, the quantum interference should be detectable for a broad range of field amplitudes $E_{THz}$ and THz photon energies. In fact, both experimental (Fig. 3a) and theoretical (Fig. 3b) traces of $I_{HH}(t)$ remain strongly unipolar and quasi-instantaneous. When the THz frequency is changed between 25 and 34 THz, the temporal separation of the emission bursts grows with the oscillation period of the driving field, but the HH maxima stay synchronized with the positive peak of the driving THz field. Both measured (Fig. 3c) and computed (Fig. 3d) traces of $I_{HH}(t)$ also remain unipolar and quasi-instantaneous when the THz field amplitude $E_{THz}$ is changed. The contrast is even enhanced for higher field strengths.

Recent studies have demonstrated that strong THz fields, needed to create HH emission, can also coherently accelerate electrons throughout the Brillouin zone before scattering occurs[4-6,21,27]. Due to resulting dynamical Bloch oscillations, electrons may undergo one or more Bragg reflections within one half-cycle of the driving field, emitting high-frequency radiation at the quasi-instantaneous Bloch frequency $\nu_B$. Since $\nu_B$ is proportional to $E_{THz}$ (see Ref. 28) the frequency of the Bloch-related contribution to HHG should trace the temporal profile of the driving field. Our XFROG algorithm allows us to retrieve the temporal phase $\phi_{HH}(t)$ of the HH pulse train (Extended Data Fig. 7), from which we obtain its instantaneous frequency $\nu_i(t) = (2\pi)^{-1}\partial\phi_{HH}/\partial t$, weighted by the spectral amplitude within our detection bandwidth (Fig. 3e). All time traces of $\nu_i(t)$ measured for different THz amplitudes follow a universal double-chirp pattern which is a fingerprint of dynamical Bloch oscillations: After a monotonic increase during the rising slope of $E_{THz}(t)$, $\nu_i$ peaks approximately at the maximum of the applied field and decreases again following the abating driving field. With increasing amplitude, the instantaneous frequency in a single HH pulse blue-shifts globally while its maximum broadens, develops shoulders and finally morphs into a non-monotonic pattern for the highest field strengths. Our quantum theory reproduces even these nontrivial features well (Fig. 3f).

The combination of non-perturbative quantum interference and dynamical Bloch oscillations may be systematically harnessed for ultrashort pulse shaping. By varying the THz carrier frequency (Figs. 3a,b) and the carrier-envelope phase of the driving waveform (Extended Data Fig. 8), the global shape of the HH pulse sequence can be tailored, whereas the frequency modulation within individual bursts is reproducibly set by the THz amplitude. Almost bandwidth-limited pulses may be generated – especially if the phase-flattening effect for high peak fields is exploited (Figs. 3e,f). We expect that in our experiment, suitable high-pass filtering of the HH pulses may allow for pulse durations as short as 3 fs in the infrared and visible domain (Extended Data Fig. 9). Since the principle of solid-based HHG is fully scalable to the UV, even shorter pulses may be possible in wide-gap materials.

In conclusion, the relative timing of HHG with respect to the driving field, the unipolar response, and a non-monotonic frequency modulation provide direct insight into the terahertz strong-field driven motion of electrons in gallium selenide. We identify a non-perturbative quantum interference between interband transitions as a salient HH generation mechanism. In its most generic form, this strong-field mechanism can occur if (i) there is a closed-loop triangle system of states that are all mutually coupled by dipole transitions (Extended Data Figure 10) and (ii) the THz field is far below these resonances and (iii) strong enough to generate non-perturbative excitations changing carrier populations significantly (see Methods and Extended Data Fig. 5). In contrast to established techniques of perturbative quantum interference between one- and two-photon transitions inducing directed charge and spin currents[29], our new concept is robust even at extremely strong fields. Thus, it may inspire new ways for quantum-logic operations[30] based on sturdy, non-perturbative transitions between strongly coupled energy bands. Driving coherences between initially fully occupied valence bands, HHG provides all-optical access even to details of the bandstructure hidden below the Fermi level. Furthermore the direct observation of lightwave controlled electron dynamics marks the way towards a complete microscopic picture of HHG in solids, ultrafast electronics, and novel solid-state CEP-stable attosecond sources.

**Acknowledgements** This work was supported by the European Research Council through grant no. 305003 (QUANTUMsubCYCLE) and the Deutsche Forschungsgemeinschaft (grant no. KI 917/2-1).

**Author contributions** M.H., F.L., O.S. and U.H. contributed equally to this work. M.H., F.L., O.S., U.H., S.W.K., M. Kira and R.H. conceived the study. M.H., F.L., O.S., M. Knorr and R.H. carried out the experiment and analysed the data. U.H., S.W.K. and M. Kira developed the quantum-mechanical model and carried out the computations. M.H., F.L., O.S., U.H., S.W.K., M. Kira and R.H. wrote the manuscript. All authors discussed the results.

**Author informations** Reprints and permissions information is available at www.nature.com/reprints. The authors declare no competing financial interests. Correspondence and requests for materials should be addressed to R.H. (rupert.huber@physik.uni-regensburg.de) or M. Kira (mackillo.kira@physik.uni-marburg.de).

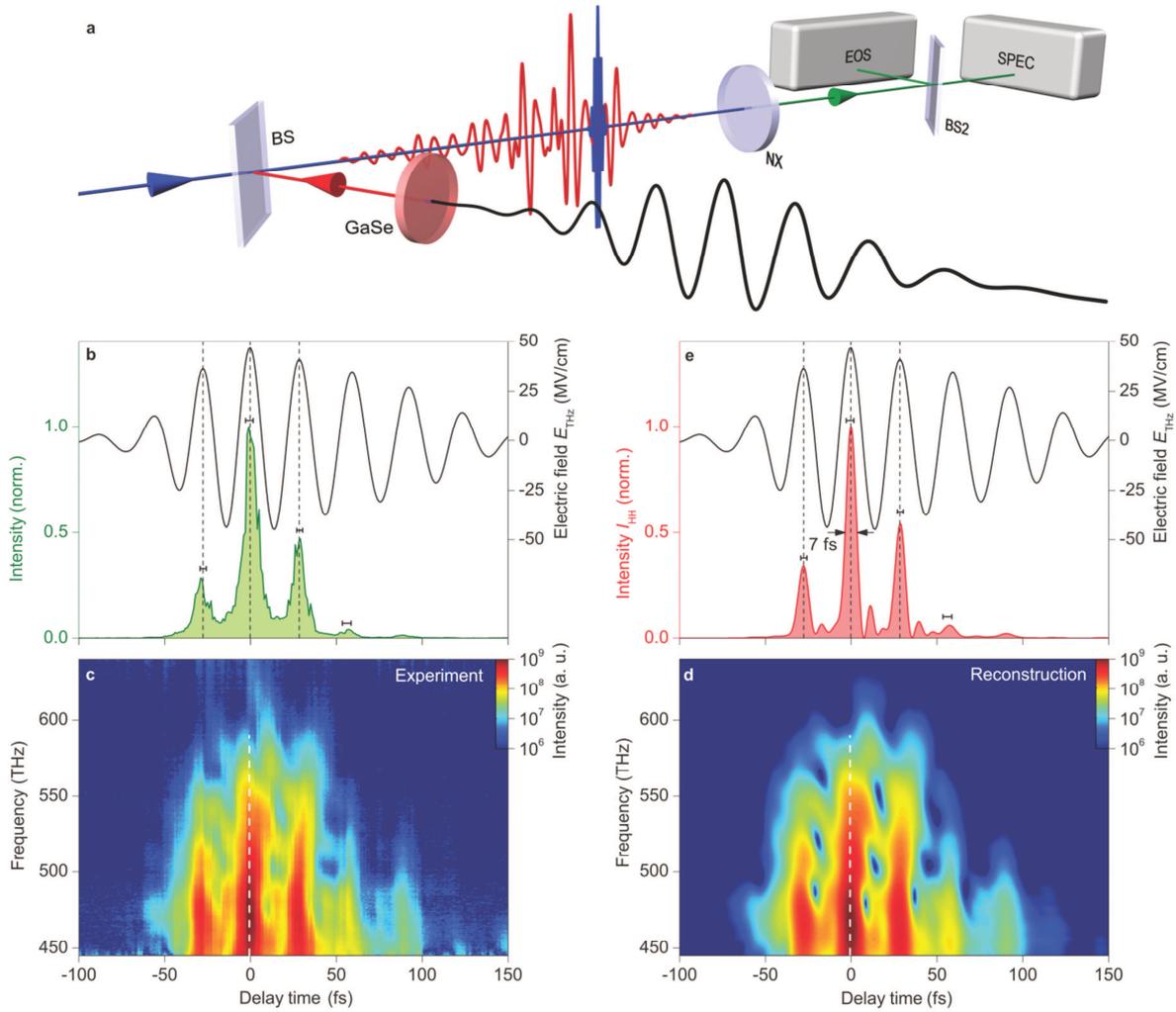

**Figure 1 | Subcycle time structure of HH emission from a bulk crystalline solid. a,** Experimental setup of the novel cross-correlation scheme: A multi-THz transient (black) is focused onto a bulk GaSe crystal (thickness: 60 µm) for HH generation. The resulting waveforms (red) are overlapped with a near-infrared gating pulse (blue, pulse duration, 8 fs, centre wavelength, 840 nm) using a beam splitter (BS) and focused into a BBO crystal (NX, thickness: 10 µm) for simultaneous electro-optic interaction and sum frequency generation. These signals are recorded with a standard electro-optic sampling (EOS) setup and a spectrograph with a cooled silicon CCD detector (Spec). **b,** Waveform of the multi-THz driving field featuring peak amplitudes of 47 MV/cm and a central frequency of 33 THz confirmed by electro-optic detection in a ZnTe crystal (thickness: 6.5 µm, black curve). Sum-frequency signals between HH and gating pulses integrated over a frequency window from 490 THz to 523 THz (green curve). Dashed vertical lines highlight the local maxima of the THz field and error bars indicate the standard deviation of the extracted SF peak position for 12 separate measurements. **c-d,** Spectrograms showing the intensity for the measured SF signal for different delay times and frequencies as recorded with a Si CCD detector (**c**) and reconstructed using a double-blind XFROG-algorithm (**d**), respectively. White dashed lines highlight the maximum SF intensity. **e,** Temporal shape of intensity $I_{HH}$ (red) of the reconstructed HH pulse sequence relative to the driving multi-THz waveform (black). Dashed lines and error bars are the same as in **b**.

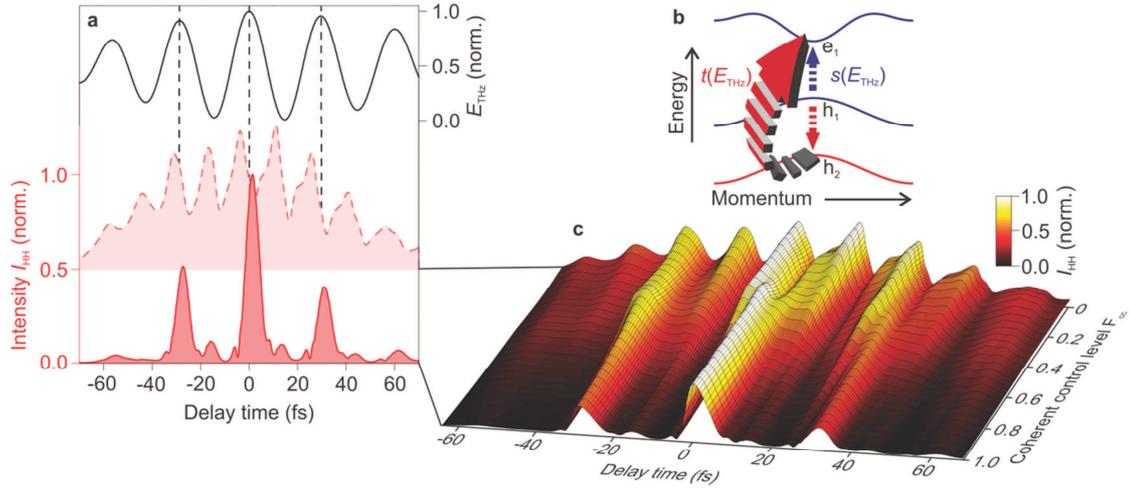

**Figure 2 | Non-perturbative quantum interference in HH emission. a,** Driving terahertz field (black curve) and calculated intensity envelope of HH emission as a function of time for $F_{cc} = 0$ (broken red curve, magnified by a factor of 25) and $F_{cc} = 1$ (solid red curve), respectively. Dashed vertical lines highlight the local maxima of the THz field. **b,** Simplified 3-band schematic of different ionization pathways from valence band $h_1$ to conduction band $e_1$. The direct transition amplitude is an odd function of the driving field $s(-E_{THz}) = -s(E_{THz})$. The amplitude $t(E_{THz})$ of the indirect path ($h_1 \rightarrow h_2 \rightarrow e_1$) is the product of two odd functions, resulting in even symmetry $t(-E_{THz}) = t(E_{THz})$. The indirect path features a transition ($h_1 \rightarrow h_2$) that is initially blocked by Pauli exclusion and only opens under strong-field excitation. **c,** High-harmonic intensity envelopes computed within the 5-band model as a function of delay time and the coherent control factor $F_{cc}$ regulating coherent transitions between occupied valence bands. Bright colours mark strong emission, dark colours mark weaker emission. All timetraces are normalised separately.

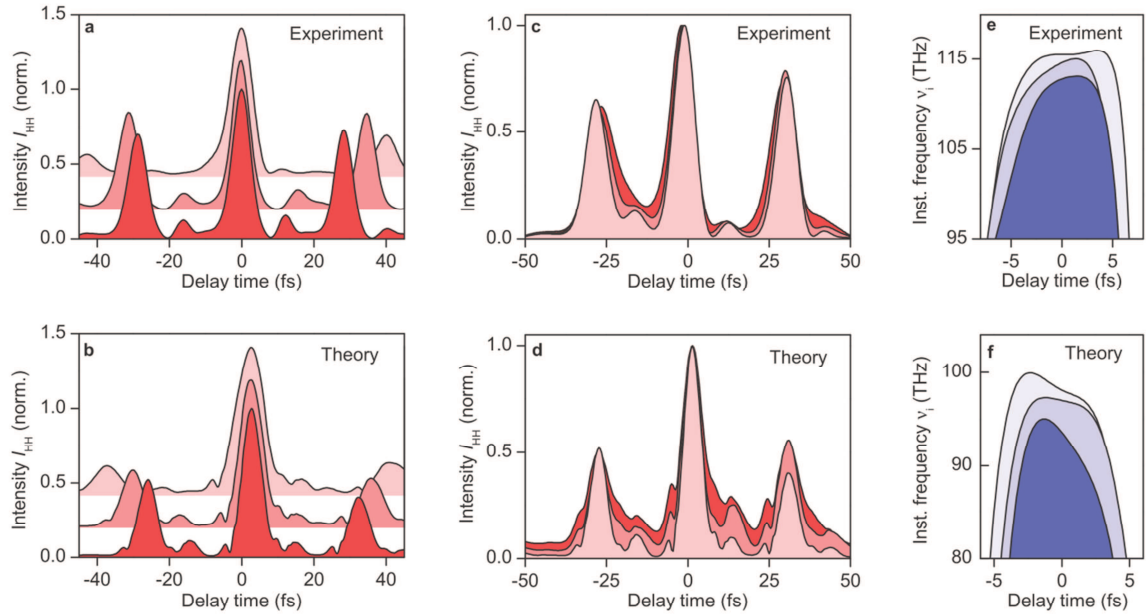

**Figure 3 | Tunability and robustness of non-perturbative quantum interference. a-b,** Measured (**a**) and calculated (**b**) intensity envelopes $I_{HH}$ of emitted HH for driving fields featuring central frequencies of 25, 30 and 34 THz, respectively. Darker colours represent higher central frequencies. **c-d,** Measured (**c**) and calculated (**d**) HH intensity envelopes for driving peak fields of 26, 31 and 44 MV/cm (experiment) and 19, 22 and 31 MV/cm (theory). Brighter colours represent higher peak fields. **e-f,** Instantaneous frequency $\nu_i$ of the central HH emission bursts shown in **c** and **d**, respectively. Brighter colours represent higher peak fields.